\documentclass[10pt,letterpaper]{article}
\usepackage{authblk}

\usepackage[utf8]{inputenc}

\usepackage{cite}

\usepackage{nameref,hyperref}

\usepackage[right]{lineno}

\usepackage{microtype}
\DisableLigatures[f]{encoding = *, family = * }

\raggedright
\usepackage{graphicx}
\usepackage{multirow}
\usepackage{geometry}
\usepackage{amsmath,amssymb}
\usepackage{array, booktabs, makecell}
\usepackage{siunitx, mhchem}
\usepackage{subfig}
\usepackage{float}

\usepackage{changepage}

\usepackage[aboveskip=1pt,labelfont=bf,labelsep=period,singlelinecheck=off]{caption}

\makeatletter
\renewcommand{\@biblabel}[1]{\quad#1.}
\makeatother

\usepackage{lastpage,fancyhdr,graphicx}
\usepackage[toc,page]{appendix}
\usepackage{listings}

\usepackage{color}

\definecolor{Gray}{gray}{.25}

\usepackage{graphicx}

\usepackage{sidecap}

\usepackage{wrapfig}
\usepackage[pscoord]{eso-pic}
\usepackage[fulladjust]{marginnote}
\reversemarginpar

\usepackage{blindtext}
\usepackage{babel}
\usepackage[most]{tcolorbox}
\usepackage{tikz}
\usepackage{varwidth}
\usepackage{capt-of}
\usepackage[export]{adjustbox}

\tcbset{enhanced,colback=cyan!5!white,colframe=cyan!75!black,fonttitle=\bfseries}

\begin{document}

\title{Low incidence rate of COVID-19 undermines confidence in estimation of the vaccine efficacy}

\author[1]{Yasin Memari \thanks{\href{mailto:ym255@cam.ac.uk}{ym255@cam.ac.uk}}}
\affil[1]{MRC Cancer Unit, University of Cambridge, Cambridge CB2 0XZ, UK}

\maketitle

\section*{Abstract}
Knowing the true effect size of clinical interventions in randomised clinical trials is key to informing the public health policies. Vaccine efficacy is defined in terms of the relative risk or the ratio of two disease risks. However, only approximate methods are available for estimating the variance of the relative risk. In this article, we show using a probabilistic model that uncertainty in the efficacy rate could be underestimated when the disease risk is low. Factoring in the baseline rate of the disease, we estimate broader confidence intervals for the efficacy rates of the vaccines recently developed for COVID-19. We propose new confidence intervals for the relative risk. We further show that sample sizes required for phase 3 efficacy trials are routinely underestimated and propose a new method for sample size calculation where the efficacy is of interest. We also discuss the deleterious effects of classification bias which is particularly relevant at low disease prevalence.


\section*{Introduction}
Vaccines are seen as the best control measure for the coronavirus pandemic. In this context, understanding the true efficacy of the vaccines and clinical interventions is crucial. Randomised clinical trials are conducted to systematically study the safety and the efficacy of an intervention in a subset of the population before it is widely used in the general population. In placebo-controlled vaccine trials, participants are randomised into vaccinated and unvaccinated groups where cases of the disease or infection are allowed to accrue over time. In planning a clinical trial, advance sample size calculation determines the size of the trial population needed to detect a minimal clinically relevant difference between the two groups if such a difference exists. The indicator for effectiveness of a vaccine is usually reduction of the cases in the vaccinated group relative to the control group. However, it is sometimes naively assumed that the trial participants who do not experience the event provide no information. Consequently, the event rate or the incidence rate of the disease receives inadequate attention. For rare diseases, it is often simply accepted that the accrual of the cases takes longer. Human clinical trials are also an area where theory and practice are seldom consistent, as experiments in human populations are hardly fully controlled experiments, not least due to unrealistic assumptions, loss to follow-up, noncompliance, heterogeneity of treatment effect and the trial population, etc. \cite{10.1093epirev} Therefore, it is not uncommon that, by the time an interim analysis declares a significant finding, the original assumptions used to define the statistical power of the study and the sample size are neglected.

In this article our interest is on evaluating the impact of the event rate, insofar as it could affect the estimation of the efficacy rate. We show that low incidence rate of the disease could lead to overestimation of confidence in the estimated efficacy rates. We propose a new method for posterior probability of the vaccine efficacy that has a more subtle relationship with the event rate. Using our approach, we obtain broader confidence intervals for the efficacy of the vaccines recently developed for COVID-19. Based on our findings, we propose new confidence intervals for the relative risk. A new method for sample size calculation in controlled efficacy trials is proposed which is more robust at low disease prevalence. Also highlighted is the impact of classification bias which could have large consequences when the disease risk is low.

\section*{Methods}
Vaccine \textit{efficacy} is defined as the proportionate reduction in the risk of disease or infection in a vaccinated group compared to an unvaccinated group. It is defined as (1-RR)$\times$100\% in terms of the relative risk or the \textit{risk ratio}, $\textrm{RR}=\pi_v/\pi_c$, where $\pi$ are the incidence of the disease among those exposed in the vaccinated and control groups. Throughout this paper we interchangeably use the terms, incidence rate, disease risk, prevalence and event rate. 

It is important to remember that the variables $\pi_v$ and $\pi_c$ are scaled binomials as they represent sample proportions. Assuming equal person-time exposure in the two groups, the efficacy is often summarised in terms of the numbers of cases in the vaccinated and unvaccinated groups, $t_v$ and $t_c$ respectively:
\begin{equation}
\alpha=1-\frac{\pi_v}{\pi_c}\simeq 1-\frac{t_v}{t_c}.\label{eqn:alpha}
\end{equation}

It appears in the literature that only approximate methods are available for the variance of the ratio of two binomial parameters \cite{pmid3291957,Katz1978OBTAININGCI}. The consensus method that is commonly used to assign confidence intervals to the risk ratio is credited to Katz et al \cite{Katz1978OBTAININGCI}. The method is based on asymptotic normality of logarithm of the ratio of two binomial variables. Assuming independence of the incidence rates, it follows that $\textrm{var}(\log(\pi_v/\pi_c))$ = $\textrm{var}(\log(\pi_v))+\textrm{var}(\log(\pi_c))$. Using a Taylor series, the variances are approximated as $\textrm{var}(\log(\pi)) \approx$ $\textrm{var}(\pi)/\pi^2$ where Wald method is often used to set $\textrm{var}(\pi)$. Then two-sided 95\% confidence intervals on the efficacy (e.g. see \cite{pmid8931208,pmid3260147,LACHENBRUCH1998569,tmi.13351}) can be written as
\begin{equation}
95\% \textrm{CL}:  1- \exp\bigg(\ln(\textrm{RR})\pm1.96\sqrt{\frac{1-\pi_v}{t_v}+\frac{1-\pi_c}{t_c}}\bigg). \label{eqn:RR}
\end{equation}
Hereafter we refer to equation \ref{eqn:RR} as pooled Wald approximation. We will show that the method underestimates the variance espcially when the incidence rate is low. 

Equation \ref{eqn:RR} sets out the large sample asymptotic variance of the risk ratio. However, Wald method used to define $\textrm{var}(\pi)$ is known to be unreliable when $\pi$ is small. One may use alternative binomial proportion confidence intervals, however, log normality of the ratio might not hold and the variance of (the logarithm of) the ratio may be irreducible. Hightower et al \cite{pmid3260147} raised question about the credibility of the confidence limits when the efficacy is high and the disease risk is low. Also, O'Neill \cite{pmid3231951} noted that, when $t\ll n$, the variance of $\ln({\textrm{RR}})$ in equation \ref{eqn:RR} remains fairly stable and quickly converges to $1/t_v+1/t_c$. 

Ratio distributions are known to have heavy tails and often no finite variance. If one were to model the likelihood function for the efficacy defined in equation \ref{eqn:alpha} in terms of independent incidence rates, the choice of the prior probabilities for $\pi_v$ and $\pi_c$ would be critical. One can readily verify that the variance of the ratio of two binomial distributions increases as binomial probabilities decrease. Uninformative priors could simply cancel out by the division and the dependence of the posterior on the prevalence would not become obvious. Analytical solutions using independent incidence rates may also be hard to obtain. 

For an analytical solution, we model the efficacy in terms of conditional probabilities of the disease risks. Independence of the probabilities of the incidence rates is neither necessary nor ideal when calculating the efficacy, as equation \ref{eqn:alpha} imposes a constraint on the two variables. Under a binomial model with overall prevalence of $\pi=t/n$ in both groups and total population size of $n$, overall number of cases $t=t_c+t_v$ follows $t\sim \mathrm{Bin}(n,\pi)$, then, from equation \ref{eqn:alpha} assuming $t_c\sim \mathrm{Bin}(t,1/(2-\alpha))$, we expect $t_c  \sim \mathrm{Bin}(n,\pi/(2-\alpha))$. Were we to use Poisson distributions for $t$ and $t_c$, $t_c$ conditional on $t$ would still follow a binomial distribution. Modeling the efficacy in terms of conditional probabilities has previously been suggested \cite{pmid8931208}. This notation enables to explicitly parametrise the likelihood function in terms of the prevalence, irrespective of the priors for $\pi_v$ and $\pi_c$.

For a general solution accounting for classification bias we assume an imperfect diagnostic procedure with sensitivity Se and specificity Sp. Then fraction of individuals who test positive for the disease is sum of true positive rate and false positive rate:
\begin{align}
T&=\textrm{Se}\times\pi+(1-\textrm{Sp})\times(1-\pi) \nonumber \\
&=c_1+c_2\pi, \label{eqn:lambda}
\end{align}
where $c_1$=1-Sp is the false positive rate and $c_2$=Se+Sp-1. The posterior distribution of $\alpha$ given that $t_c$ is binomial follows as 
\begin{align}
p(\alpha | t_c,\pi,c_1,c_2)&=\frac{p(t_c | \alpha,\pi,c_1,c_2)p(\pi)p(\alpha)}{g(\alpha)} \nonumber \\
&\propto\frac{1}{g(\alpha)}\binom{n}{t_c}\left(\frac{c_1+c_2\pi}{2-\alpha}\right)^{t_c}\left(1-\frac{c_1+c_2\pi}{2-\alpha}\right)^{n-t_c}f(\pi), \label{eqn:post}
\end{align}
where $f(\pi)$ is the prior on $\pi$ and we have assumed uniform prior on the efficacy $\alpha\sim \mathrm{unif}\{0,1\}$. For a complete solution, the marginal likelihood $g(\alpha)$ can be written in terms of the incomplete beta function (see e.g. \cite{608719}):
\begin{align*}
g(\alpha)&=f(\pi) \binom{n}{t_c} (c_1+c_2\pi) \big\{B(c_1+c_2\pi;t_c-1,n-t_c+1) \\
&- B((c_1+c_2\pi)/2;t_c-1,n-t_c+1)\big\}.
\end{align*}
As we do not intend to impose a prior on the prevalence, $f(\pi)$ in equation \ref{eqn:post} cancels out and our analysis, in essence, is likelihood based. One needs to remember that, the posterior in equation \ref{eqn:post}, as it was derived from the second equality in equation \ref{eqn:alpha}, is valid only when the individuals are equally divided between the two groups. 

The mode of the posterior of $\alpha$ is obtained by setting the derivative of the log likelihood to zero i.e. $\partial \ell/\partial\alpha=\partial \mathrm{ln}(p(\alpha | t_c,\pi))/\partial \alpha=0$. This leads to 
\begin{equation}
\alpha_{mode}=2-\frac{n(c_1+c_2\pi)}{t_c} \label{eqn:mode},
\end{equation}
which corresponds to the maximum likelihood estimator (MLE). Cramér–Rao bound expresses a lower bound on the variance of any unbiased estimator of $\alpha$ in terms of the inverse of the Fisher information
\begin{equation}
\textrm{Var}(\alpha_{mode})\ge\frac{1}{\mathcal{I}(\alpha)}, \label{eqn:cramer}
\end{equation}
where the Fisher information $\mathcal{I(\alpha)}$ is obtained as
\begin{align}
\mathcal{I}(\alpha)= \mathbb{E}\Big[\Big(\frac{\partial}{\partial\alpha}\ell(\alpha | t_c,\pi)\Big)^2\Big] &=n \times  \mathbb{E}\Big[\Big(-\frac{1-t_c}{2-\alpha-\pi}-\frac{-1}{2-\alpha}\Big)^2\Big] \nonumber \\
&= \frac{n (c_1+c_2\pi)}{(2-\alpha)^2(2-\alpha-(c_1+c_2\pi))}. \label{eqn:fisher}
\end{align}
Here $\mathbb{E}$ denotes `expected' over $t_c$, where we have substituted $\mathbb{E}[t_c]=\mathbb{E}[t_c^2]=(c_1+c_2\pi)/(2-\alpha)$. We will show that the conditional binomial model has a more subtle dependence on $\pi$ compared to the pooled Wald method. 

Under certain regularity conditions and assuming asymptotic normality near MLE, 95\% confidence intervals on $\alpha_{mode}$ can be estimated as
\begin{equation}
\alpha_{mode}\pm\frac{1.96}{\sqrt{\mathcal{I}(\alpha_{mode})}}. \label{eqn:cramerCIs}
\end{equation}
However, as the posterior distribution is asymmetric, especially when the efficacy is high, and the intervals could lie outside [0,1], we will estimate the credible intervals computationally.

\section*{Results}

\subsection*{Effect of incidence rate on vaccine efficacy}

The posterior probability of vaccine efficacy given in its simplest form in equation \ref{eqn:post} is ready for inspection. Using binomial notation is particularly useful in enabling us to directly plug in the numbers $n$, $t_c$ in the estimation of $\alpha$. In this section we evaluate the impact of the incidence rate on the efficacy and assign new confidence bounds to the efficacy of COVID-19 vaccines.

\begin{figure}[!t]
 \centering
    \mbox{\includegraphics[width=2.7in]{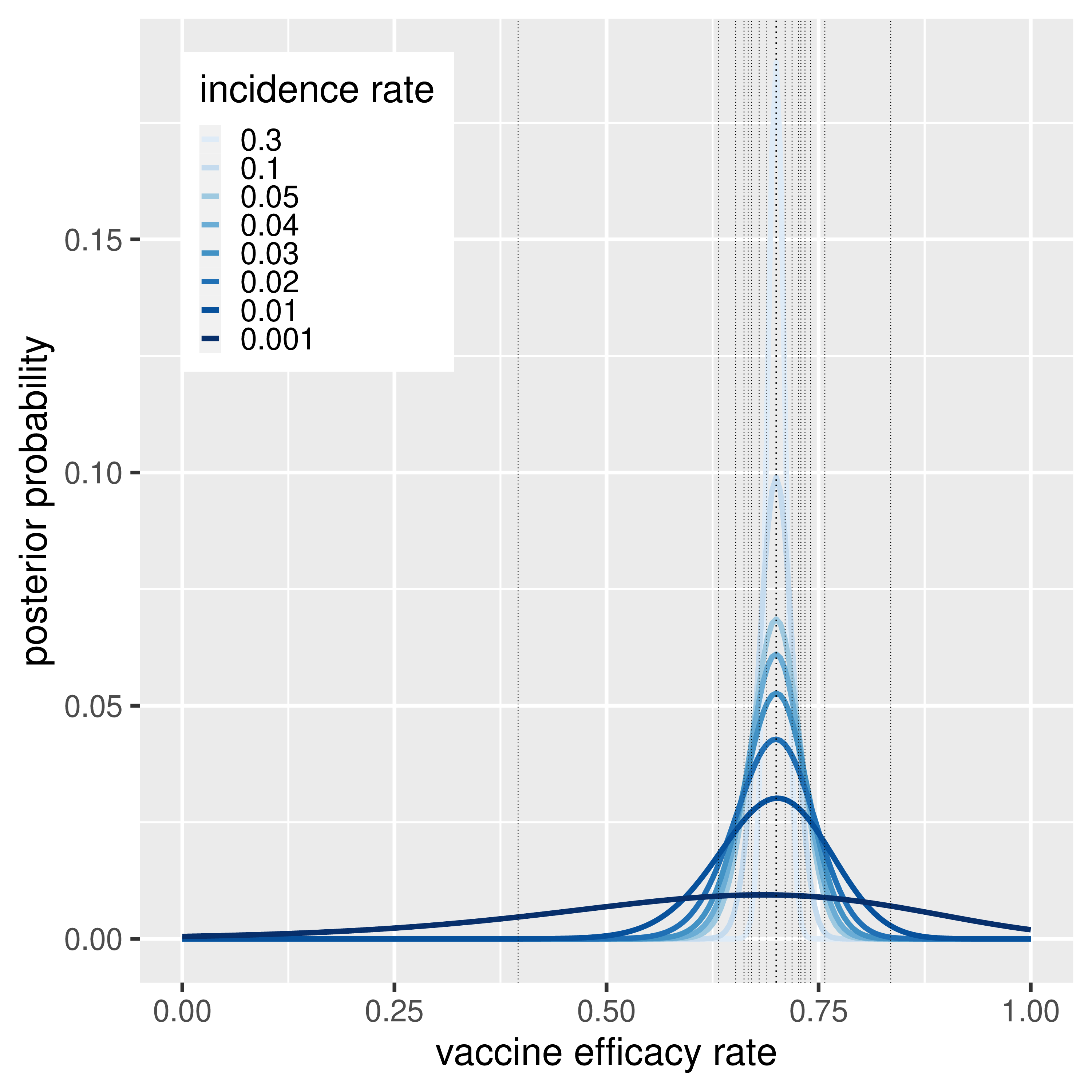}}   
    \hspace{1px}
    \mbox{\includegraphics[width=2.7in]{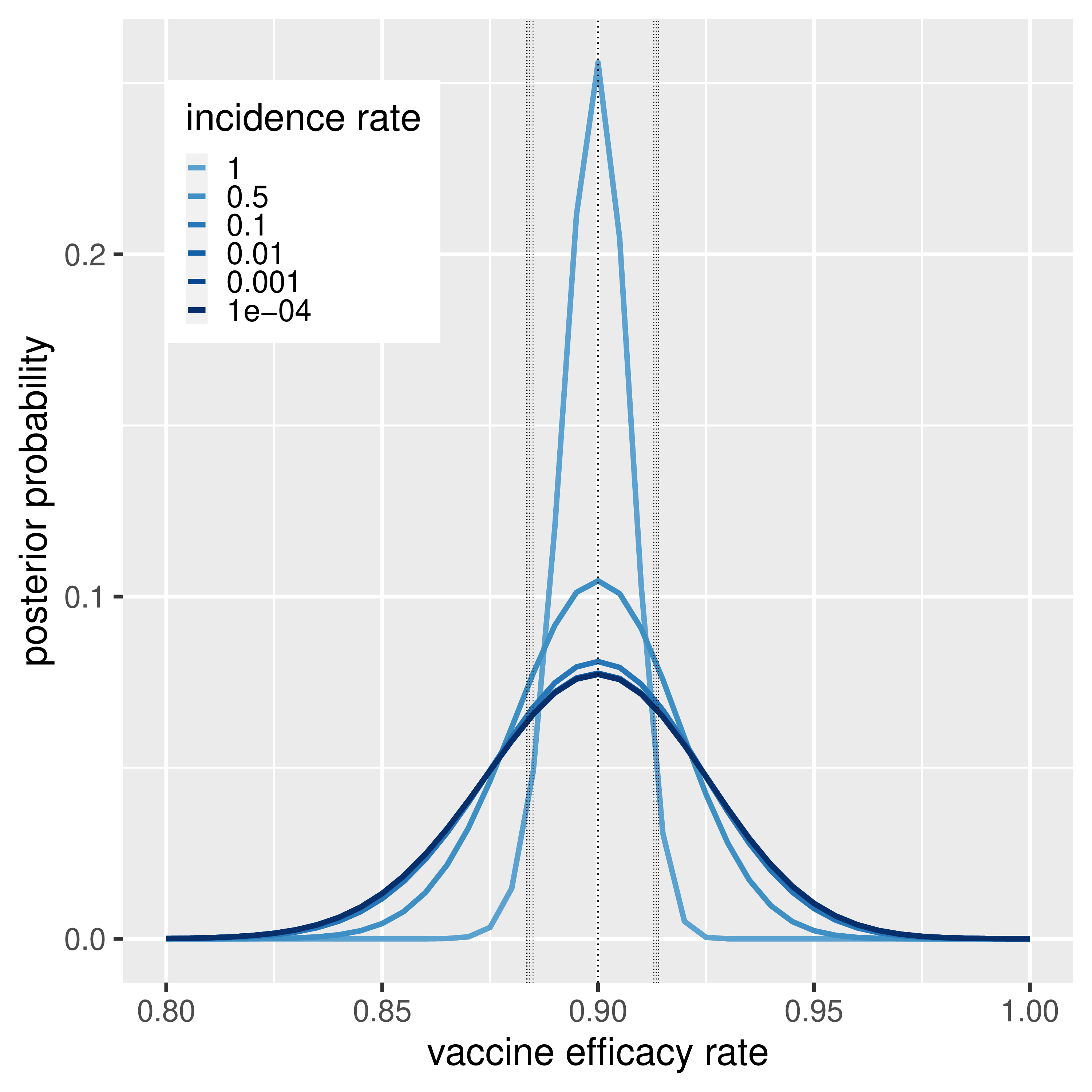}}
	\caption{\color{Gray} \textbf{Posterior distribution of vaccine efficacy}. Blue lines represent the normalised posterior probabilities, while vertical lines show the independent pooled Wald confidence intervals. Left hand plot assumes a fixed $n$=50,000 while right hand plot is for a fixed $t$=2,000. The general trend holds for different values of the parameters. Wald method overstates the confidence in the efficacy when $t\ll n$.}
    \label{fig:figure-1}
\end{figure}

Firstly, we assume a diagnostic test with perfect sensitivity and specificity i.e. Se=Sp=1. In the absence of misclassification, mode of the posterior in equation \ref{eqn:mode} corresponds to the expectation $\hat{\alpha}=1-t_v/t_c$. The larger $n$ the smaller the variance of the posterior, however, for a fixed $n$, the variance depends on $\pi$. Figure \ref{fig:figure-1} shows the posterior probability of $\alpha$ plotted over a range of $\pi$, for a fixed $n$ on the left hand, and for a fixed $t$ on the right hand, assuming true vaccine efficacy of 70\% and 90\% respectively. Also plotted in vertical lines are the independent 95\% confidence intervals from equation \ref{eqn:RR}.  As the event rate falls, the posterior distributions and the confidence intervals become wider, however, for a fixed $t$ (right plot) Wald intervals are stable over a wide range of $\pi$, and more so when the efficacy is high. The proposed conditional binomial model better represents the variability at low prevalence.

Three clinical trials of the vaccines designed to prevent COVID-19 recently published their interim phase 3 analysis results \cite{pmid33306989,NEJMoa2034577,NEJMoa2035389}  with two of them reporting incredibly narrow 95\% confidence bounds on the efficacy. The reported case numbers and the efficacy rates for the primary end points are provided in Table \ref{tab:table-1}. Firstly, we note that, although the trials used different models and priors on the efficacy, the reported confidence intervals almost perfectly correspond with those obtained from equation \ref{eqn:RR}. At large $n$ the posterior is clearly dominated by the data and the Bayesian and the frequentist are equivalent. Furthermore, especially where the efficacy is high, pooled Wald confidence intervals hardly vary by the choice of $n$. If one were to use different values for $n$ in Table \ref{tab:table-1}, over a large range of the values equation \ref{eqn:RR} would still give the same confidence intervals. Therefore, the uncertainty caused by $n_v$ and $n_c$ is not accounted for. 

\begin{table}[!b]
\caption{Estimated efficacy of COVID-19 vaccine trials}
    \label{tab:table-1}
\begin{tabular}{c@{\qquad}ccc@{\qquad}c}
  \toprule
  \multirow{2}{*}{\raisebox{-\heavyrulewidth}{Trial}} & \multicolumn{3}{c}{Case numbers and reported efficacy rates} & \multicolumn{1}{c}{Estimated efficacy rate} \\
  \cmidrule{2-5}
 & \thead{case rate in\\ vaccinated} &  \thead{case rate in\\ control} &  \thead{reported\\ efficacy and 95\% CI} &  \thead{estimated mode\\ and 95\% credible interval} \\
  \midrule
AZ-Oxford (combined) & 30/5,807 & 101/5,829 & 70·4\% [54·8, 80·6] & 70.3\% [39.1, 90.9] \\
Pfizer-BioNTech & 8/18,198 &162/18,325 & 95.0\% [90.3, 97.6] & 95.1\% [74.9, 99.6] \\
Moderna-NIH & 11/14,134 & 185/14,073 & 94.1\% [89.3, 96.8] & 94.1\% [75.4, 99.5] \\
  \bottomrule
\end{tabular}
\end{table}

We re-estimate the confidence intervals using the conditional binomial model presented in the Methods. Using the case numbers reported, the likelihood of the data in equation \ref{eqn:post} is obtained by setting the prevalence to $\pi=T=t/n$. Then maximum \textit{a posteriori} (MAP) and 95\% credible intervals for the efficacy rates are calculated computationally. The results shown in Table \ref{tab:table-1} are contrasted with those reported. Although estimated modes are the same, our credible intervals are wider. Incorporating the incidence rates has removed the overwhelming confidence originally assigned to the point estimates. Note that, our approach requires the trial participants to be equally divided between the vaccinated and unvaccinated groups which is roughly the case here. 

\begin{figure}[!t]
 \centering
    \mbox{\includegraphics[width=1.9in]{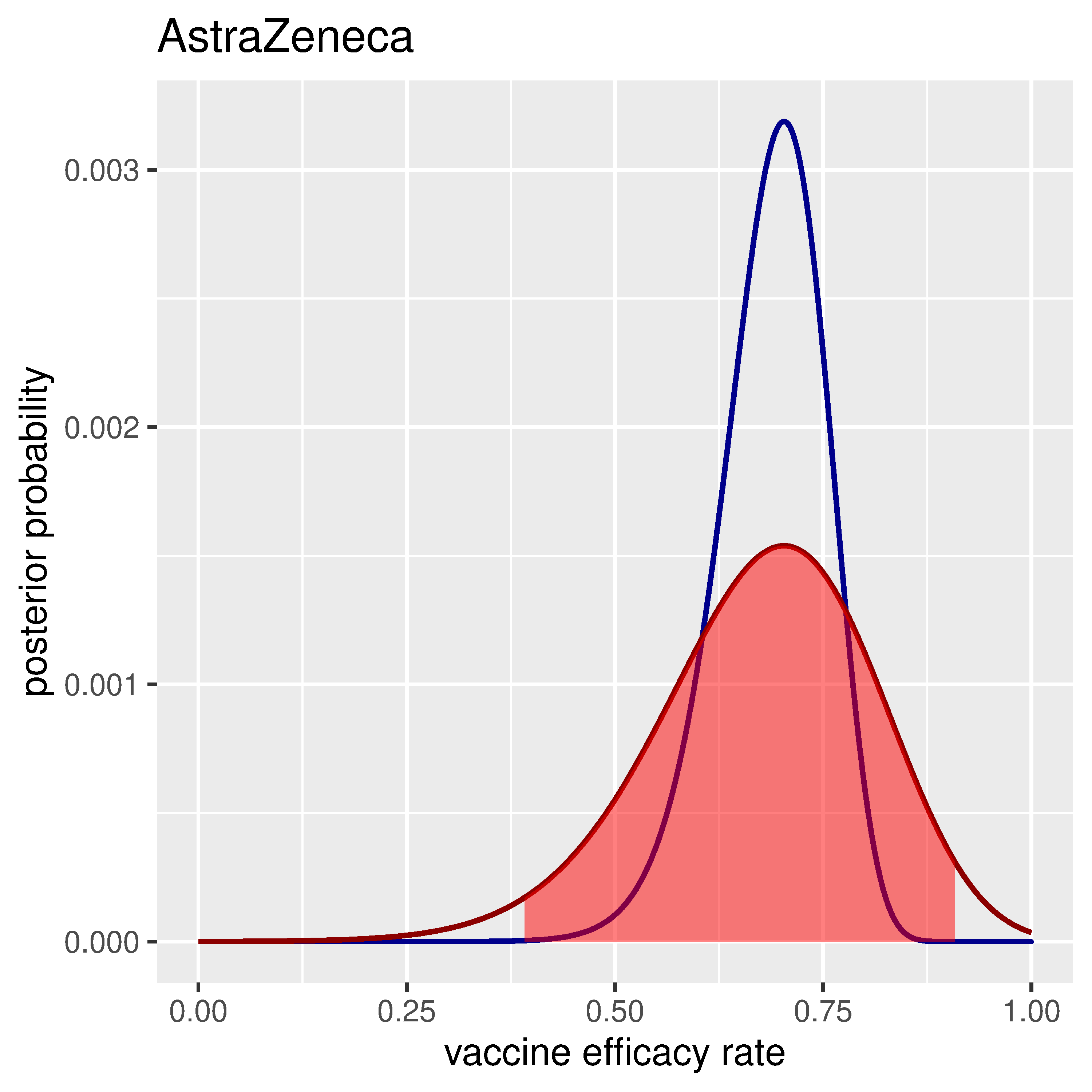}}   
    \hspace{1px}
    \mbox{\includegraphics[width=1.9in]{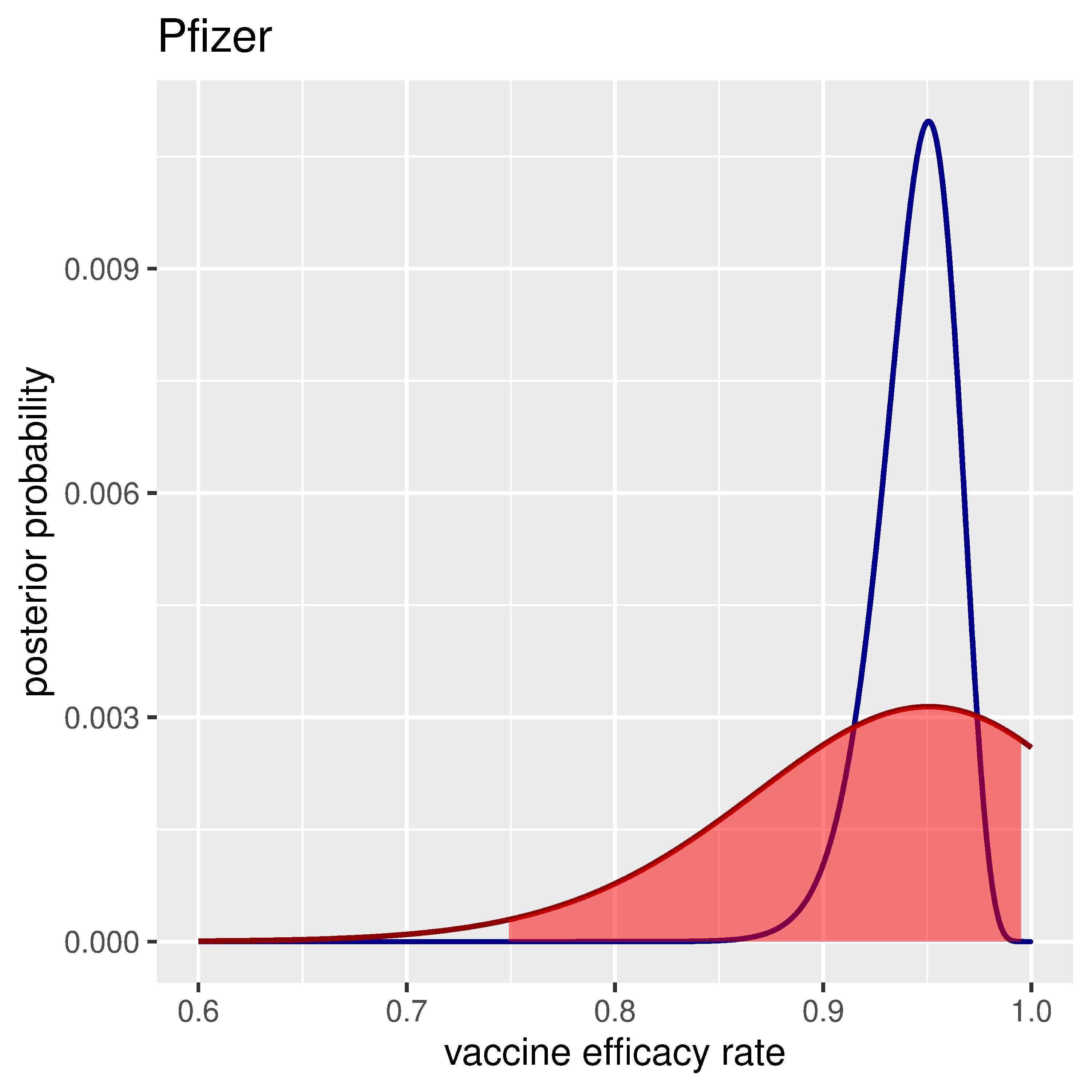}}
     \hspace{1px}
    \mbox{\includegraphics[width=1.9in]{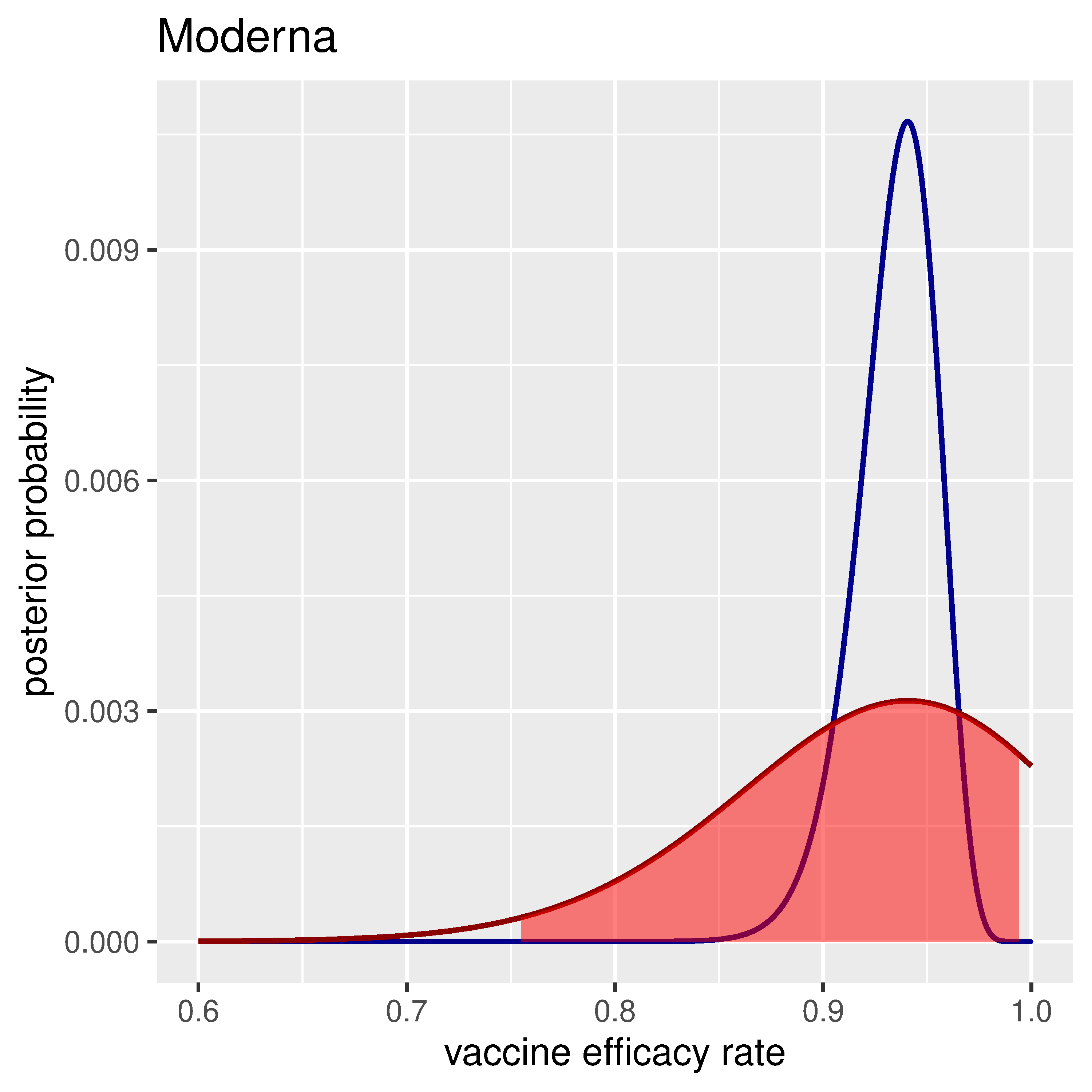}}
	\caption{\color{Gray} \textbf{Estimated efficacy of COVID-19 vaccines}. Posterior probabilities for the conditional binomial model are plotted in red, with shaded areas representing the 95\% credible intervals. Blue curves are for when $n$ is set to $t_v+t_c$ and correspond with pooled Wald approximation.}
    \label{fig:post}
\end{figure}

Figure \ref{fig:post}, in red, shows the posterior probabilities and the credible intervals for COVID-19 vaccines. Of note is that, if we were to hypothetically assume $\pi=t/n=1$, the posterior in equation \ref{eqn:post} would produce the same intervals as those reported by the vaccine trials and Wald approximation. Moreover, an independent binomial model with uninformative (e.g. uniform) priors for $\pi_v$ and $\pi_c$ would produce the pooled Wald intervals. 

\subsection*{Bias in case classification}
So far we have assumed no bias in classification of the cases, however, imperfect diagnostic procedure could lead to misclassification of the infected and uninfected individuals. In this section we examine the effect of classification bias on estimation of the efficacy.

It is worth noting that equation \ref{eqn:lambda} requires the observed infection rate $T$  to be greater than the false positive rate $c_1=1-Sp$. This relates to the `false positive paradox' which implies that the accuracy of a diagnostic test is compromised if the test is used in a population where the incidence of the disease is lower than the false positive rate of the test itself. Furthermore, false negatives could dominate at low incidence rates. When the disease risk is low, as the majority of the tests are negative, a small false negative rate could lead to a situation where false negatives outnumber the positive cases. These concepts are further explained in Note 1.

\begin{figure}[!t]
 \centering
    \mbox{\includegraphics[width=2.7in]{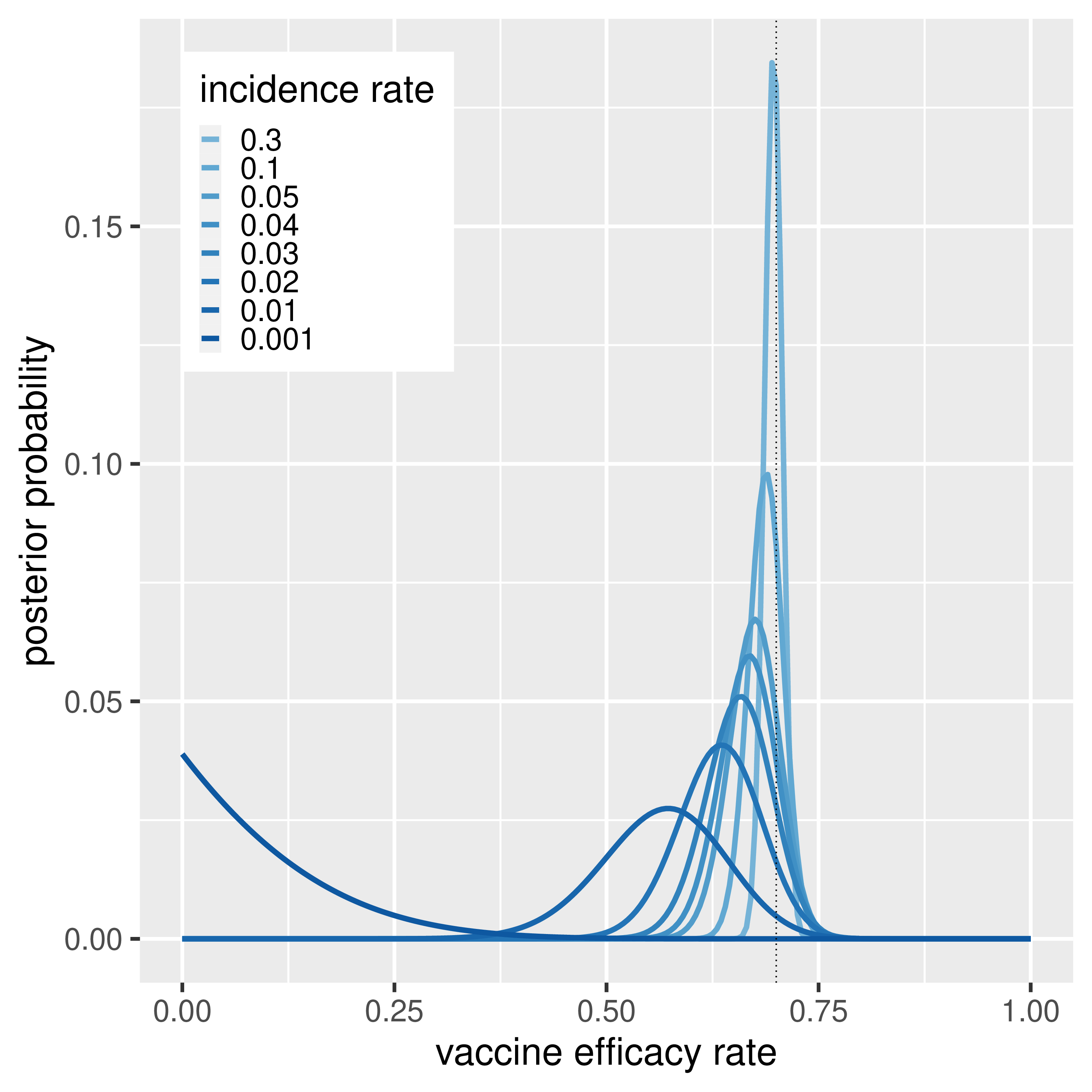}}   
    \hspace{1px}
    \mbox{\includegraphics[width=2.7in]{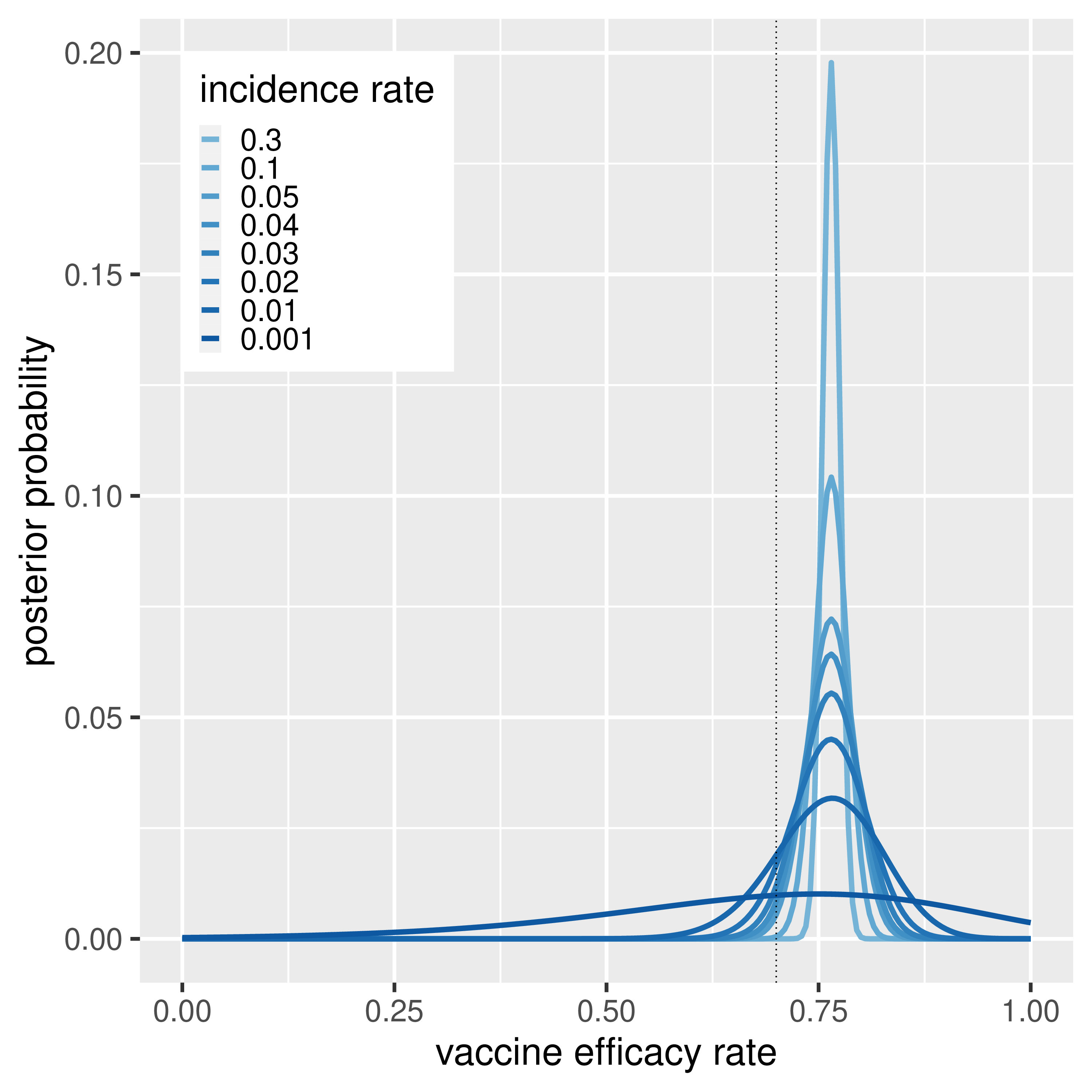}}
	\caption{\color{Gray} \textbf{Effect of imperfect diagnostic procedure}. Misclassification error biases the vaccine efficacy rate. Left plot shows the distributions for Se=1 and Sp=0.999, while the right plot is for Se=0.95 and Sp=1, with $n$=50,000 in both. True efficacy rate is assumed at 70\%. Imperfect specificity, however small, could have disastrous effects when incidence rate is low, whereas lack of sensitivity consistently inflates the efficacy rate.}
    \label{fig:figure-3}
\end{figure}

Figure \ref{fig:figure-3} illustrates the effect of classification bias on the posterior probability of the vaccine efficacy. The left plot shows the impact of a very small reduction in specificity to 0.999 (or increase in false positive rate), while the right hand plot shows the effect of reduction in sensitivity to 0.95 (or increase in false negative rate). A small loss of specificity could lead to serious underestimation of the effect size as noted by \cite{LACHENBRUCH1998569,tmi.13351}, but it could further lead to complete loss of precision when the incidence rate is low. Loss of sensitivity results in overestimation of the efficacy irrespective of the disease rate. In these plots, we have considered a larger reduction in sensitivity, not only because reduction in specificity has a more dramatic effect, but also as diagnostic assays typically have relatively higher specificity than sensitivity, not least due to specimen collection, insufficient viral load, stage of the disease, etc. \cite{pmid33301459} However, the effect of loss of sensitivity is consistently toward shifting the mode in equation \ref{eqn:mode}, or MAP, to higher values of $\alpha$, even at low incidence rates where negative predictive value is high.

\begin{tcolorbox}[float, drop shadow, title=Note 1,sidebyside,sidebyside align=top,lower separated=false]
J. Balayla \cite{pmid33027310} noted that there exists a \textit{prevalence threshold} below which the positive predictive value (PPV) of a diagnostic test drops precipitously relative to the prevalence. This means that at too low a prevalence a positive test result could more likely be a false positive than a true positive. More underappreciated is the impact of the negative predictive value (NPV). Though, at low incidence rates, the negative predictive value is nearly 100\%, a small loss in sensitivity could still have a marked effect as the negative tests vastly outnumber the positive tests. We could even have a situation where the false negatives are more than the true and false positives. To avoid these pitfalls, the participants are pre-selected for their symptoms before confirmation with the assay. Though this raises the pre-test probability, it could cause collider bias \cite{pmid33184277}.
    \tcblower
        \includegraphics[height=.3\textheight,width=\linewidth,valign=t]{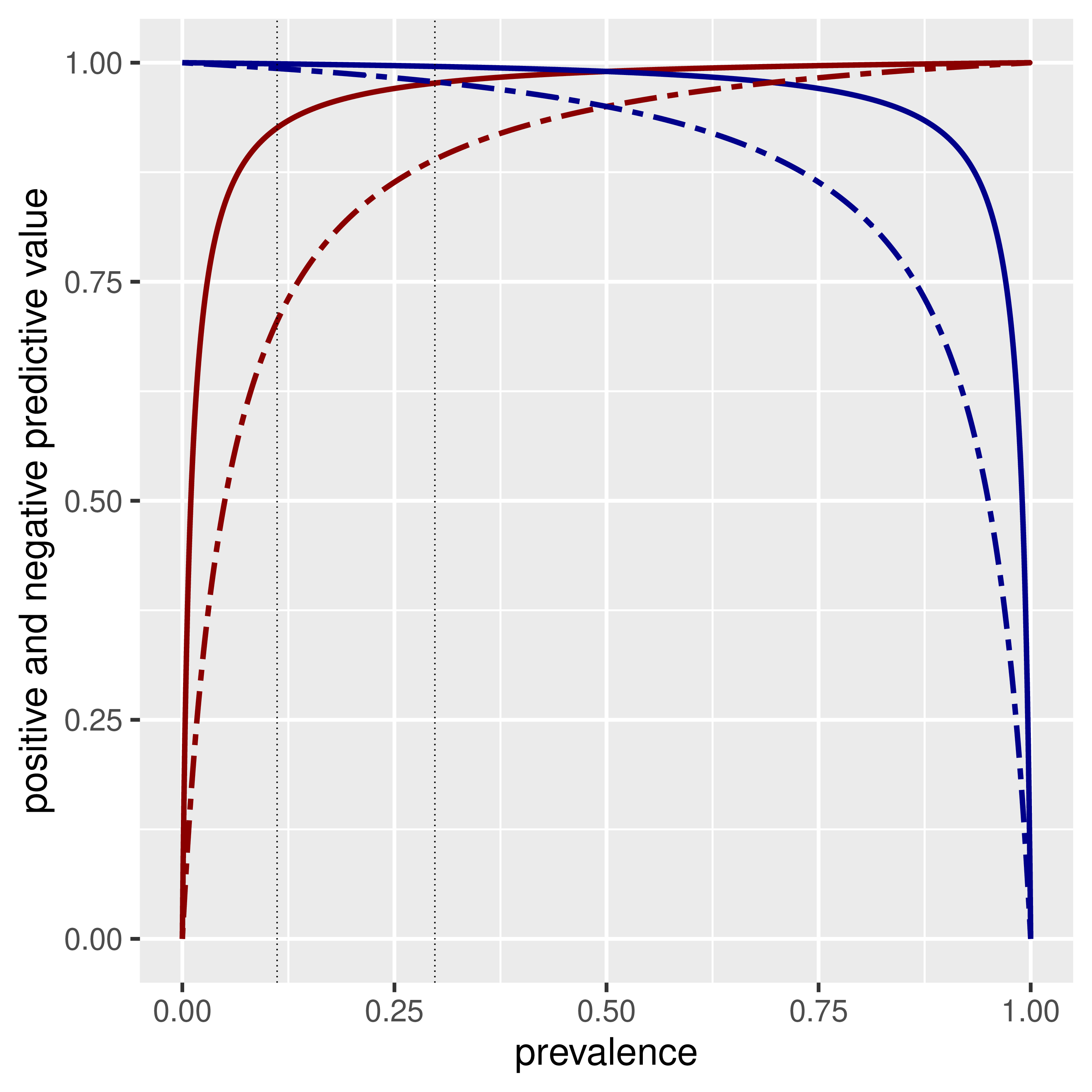}
        \captionof{figure}{Positive (red) and negative (blue) predictive values are plotted in terms of population prevalence. Solid lines are for to a diagnostic test with Se=Sp=0.99; dashed lines are for Se=Sp=0.95. Vertical lines show the prevalence thresholds.}\label{fig:ppv}
\end{tcolorbox}

\section*{Discussion}
Base rate fallacy happens in situations where base rate information is ignored in favour of individuating information. In probability terms, it often occurs when $P(A|B)$ is confused or interchangeably used with $P(B|A)$ ignoring the prior probability $P(A)$, e.g. probability of having a rare disease given a positive test is wrongly equated to probability of a positive test given the disease (or diagnostic sensitivity) ignoring the low prior probability of the disease itself. We showed, in estimation of the vaccine efficacy when the disease rate is low, not only diagnostic error could have deleterious effects, but also failure to appropriately integrate the information about the base rate or incidence rate of the disease in the calculation could lead to underestimation of the uncertainty.

Vaccine efficacy is defined in terms of the risk ratio $\pi_v/\pi_c$, that is the ratio of two binomial proportions. Ratio distributions are known to have undefined variances, conversely, pooled Wald method has been traditionally used to approximate the variance of the risk ratio. In this article, we used a parametrisation that makes the dependence of the efficacy on the disease prevalence explicit, without recourse to priors for $\pi_v$ and $\pi_c$. Particularly, improper priors for $\pi_v$ and $\pi_c$ could lead to underestimation of the variance. We conditioned $t_c$ on $t=t_c+t_v$ and treated $t$ as another random variable. The resulting compound probability $t_c \sim \mathrm{Bin}(n,\pi/(2-\alpha))$ is over-dispersed and better captures the variability of the variance with $\pi$, whereas pooled Wald confidence intervals are largely insensitive to $\pi$ when $\pi$ is small. 

Wald method is intended as large sample approximation, however, the bulk of the life sciences deals with small sample sizes. Therefore, it is likely that the confidence intervals reported in the literature for the risk ratio (and odds ratio) are overly optimistic. By analogy of equations \ref{eqn:cramer} and \ref{eqn:cramerCIs}, one could define new confidence intervals for the risk ratio by substituting RR=$(n_c/n_v)(1-\alpha)$ for unequal sized groups in the Fisher information. The results can be written as
\begin{equation}
95\% \textrm{CL}:  \textrm{RR}\pm 1.96 \frac{n_c}{n_v}\big(1+\frac{t_v}{t_c}\big) \sqrt{\frac{1+t_v/t_c-\pi}{t_v+t_c}}, \label{eqn:fisherCIs}
\end{equation}
where $\pi$=$(t_v+t_c)/(n_v+n_c)$. The above intervals on the risk ratio are generally wider than but converge to the pooled Wald method when the sample size is large. They may be preferred to those obtained from equation \ref{eqn:RR} when the sample size is small or the relative risk is low. Particularly, for a fixed sample size as RR nears zero, the upper bound in equation \ref{eqn:fisherCIs} remains conservative and the lower bound takes negative values and becomes undetermined. On the contrary, as RR nears zero, the pooled Wald intervals remain positive and shrink rapidly, giving the counterintuitive impression of increased precision when the incidence rate is low (similar to figure \ref{fig:post}). However, as with Wald method, the confidence intervals in equation \ref{eqn:fisherCIs} were derived using normal approximation which may not hold when RR significantly deviates from 1. 

Our findings have implications for pre-planning the sample sizes for phase 3 efficacy trials. Sample size calculation in case-control design is often stated as ``How many samples are needed to be randomised in order to conclude with 100$(1-\beta)$\%  power that a treatment difference of size $\Delta$ exists between the two groups at the level of significance of $\alpha$?". Therefore calculation of sample size requires specification of the null hypothesis (expected treatment effect) and the alternative hypothesis defined in terms of the difference in treatment outcomes. Here, $\alpha$ or type I error is the probability of rejecting the null hypothesis where we should not, and $\beta$ or type II error is the probability of failing to reject the null hypothesis where we should reject it. Under the assumption of normality of the treatment outcome, a generic formula for per-group sample size is derived in terms of the two-sample t-test: \cite{10.1093epirev}\begin{equation}
n=\frac{2\sigma^2}{\Delta^2}(z_{1-\alpha/2}+z_{1-\beta})^2, \label{eqn:ssize}
\end{equation}
where $z$-scores determine the critical values for the standard normal distribution. Therefore one needs to specify the variance of the measured variable, the desired rates of error and the magnitude of the treatment difference. Where the measured variable is binary (infected or uninfected), the test statistic reduces to the test for the difference between two proportions. Where the efficacy is of interest, the log normal approximation of the risk ratio from equation \ref{eqn:RR} may be used to define the test statistic. O'Neill \cite{pmid3231951} calculated the required sample sizes for a two-sided test given the pooled Wald variance in equation \ref{eqn:RR}. We re-write the \textit{total} sample size in this form: 
\begin{equation}
n=2\frac{(z_{1-\alpha/2}+z_{1-\beta})^2}{d^2}\Big(\frac{(2-\textrm{VE})^2}{\pi(1-\textrm{VE})}-2\Big), \label{eqn:ssizeWald}
\end{equation}
where
\begin{equation*}
d=\ln\Big(\Delta/(2(1-\textrm{VE}))+\sqrt{\big(\Delta/(2(1-\textrm{VE}))\big)^2+1}\Big).
\end{equation*}
Here VE is the anticipated efficacy and $\Delta$ is the expected difference in VE in absolute terms. We showed, however, that at low prevalence rate, equation \ref{eqn:RR} significantly underestimates the variance. Using an inadequately small variance could lead to underestimation of the type I and type II errors, potentially resulting in winner's curse in underpowered studies \cite{pmid18633328,pmid23571845}. If instead we were to use the proposed compound binomial model, one could simply substitute the variance in equation \ref{eqn:cramer}. As in \cite{pmid3231951}, under the assumption of normality and assuming $\Delta$ is the difference between the upper and lower limits of the confidence interval, substituting the margin of error as $\Delta/2=z\sigma$ in equation \ref{eqn:cramer} gives
\begin{equation}
n\ge 4 \frac{(z_{1-\alpha/2}+z_{1-\beta})^2}{\pi \Delta^2} (2-\textrm{VE})^2(2-\textrm{VE}-\pi). \label{eqn:ssizeCramer}
\end{equation}
This equation sets out the \textit{total} required sample size for a perfect diagnostic test, to be equally divided between the two groups. 

\begin{figure}[t]
 \centering
    \mbox{\includegraphics[width=2.7in]{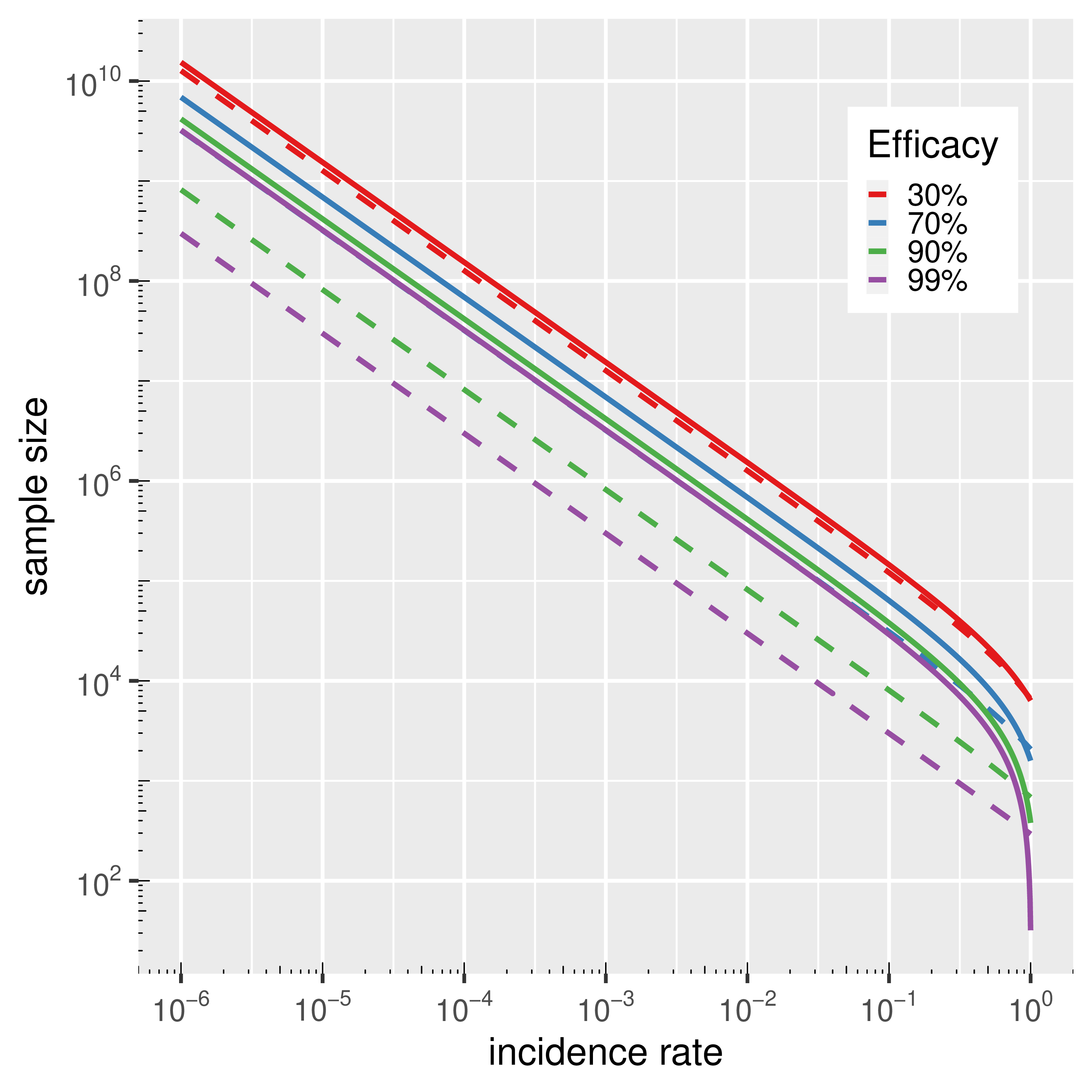}}   
    \hspace{1px}
    \mbox{\includegraphics[width=2.7in]{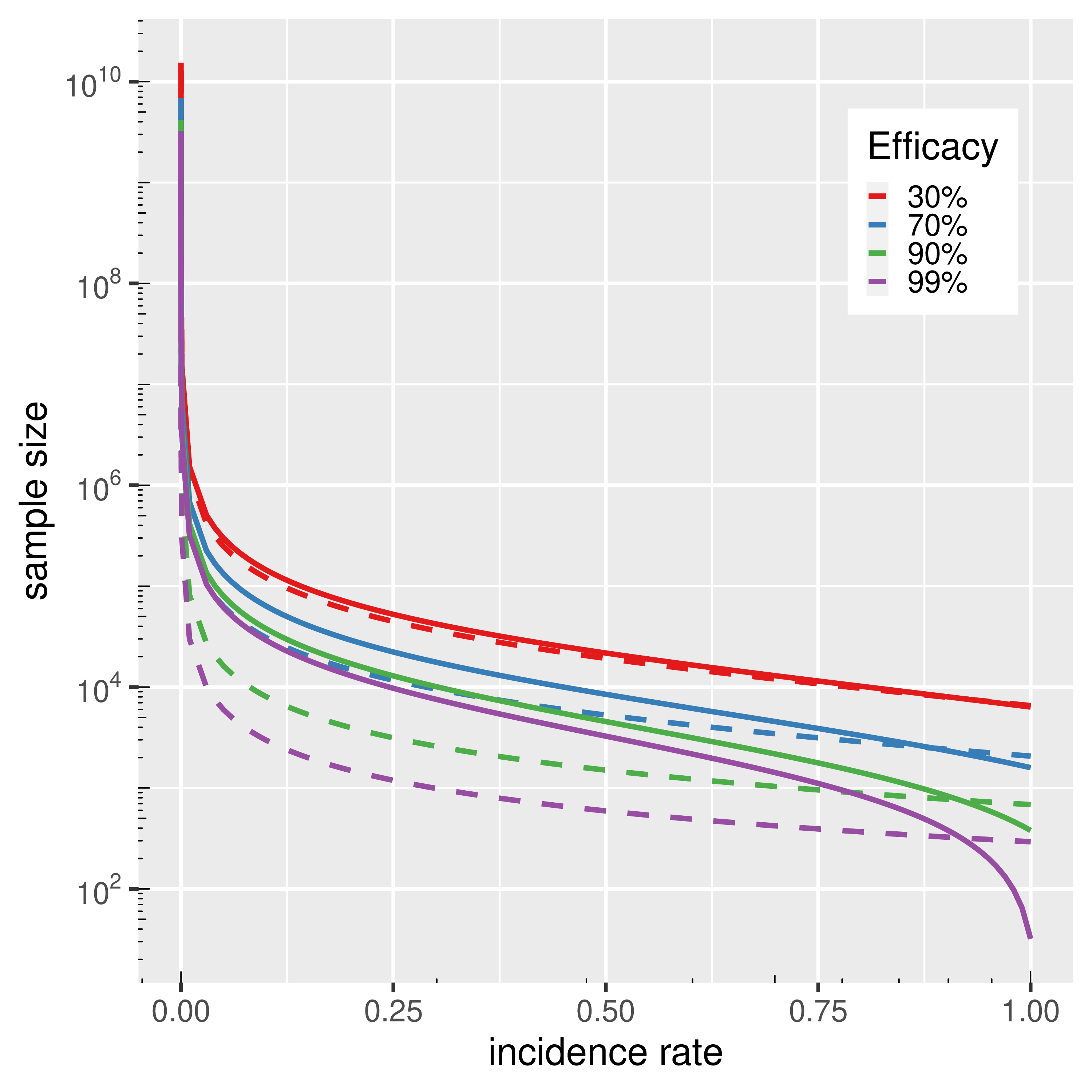}}
	\caption{\color{Gray} \textbf{Sample size relative to disease prevalence}. Total number of samples required to detect with 80\% power and level of significance of $\alpha=0.05$ a difference in the efficacy of size $\Delta=10\%$. Solid lines represent Cramér–Rao bound and dashed lines represent pooled Wald approximation. On the left, x-axis is on logarithmic scale. y-axis is logarithmic in both plots.}
    \label{fig:figure-4}
\end{figure}

The proposed Cramér–Rao bound based formula \ref{eqn:ssizeCramer} assumes normality of distributions of the null and the alternative hypotheses, however, the binomial likelihood function is asymmetric, as is pooled Wald intervals (see \cite{pmid3231951}), and becomes more so as the efficacy increases. Notwithstanding the limitations, we plug in the critical values for $\alpha=0.05$ and power of $100(1-\beta)=80$ per cent ($z_{1-\alpha/2}=1.96$ and $z_{1-\beta}=0.84$) in equations \ref{eqn:ssizeWald} and \ref{eqn:ssizeCramer}. The resulting sample sizes are plotted in Figure \ref{fig:figure-4} for $\Delta=10\%$ and different prevalence and efficacy rates. 

\begin{table}[t]
\caption{Total sample sizes needed to conclude with 80\% power and $\alpha$=0.05 a significant effect size}
    \label{tab:table-2}
\begin{tabular}{cc@{\qquad}ccccccc}
  \toprule
  \multicolumn{2}{c}{effect size} & \multicolumn{7}{c}{event rate} \\
     \cmidrule{1-2} \cmidrule{3-9}
VE & $\Delta$ & 0.5 & 0.1 & 0.05 & 0.01 & 0.005 & 0.001 & 0.0005 \\
\midrule
 0\% & 10\% &37,632 & 238,336 & 489,216 & 2,496,256 & 5,005,056 & 25,075,456 & 50,163,456 \\
 0\% & 20\% &9,408 &  59,584 & 122,304 &   624,064 & 1,251,264 &  6,268,864 & 12,540,864 \\
 0\% & 30\% &4,181 &  26,482 &  54,357 &   277,362 &   556,117 &  2,786,162 &  5,573,717 \\
 0\% & 40\% &2,352 &  14,896 &  30,576 &   156,016 &   312,816 &  1,567,216 &  3,135,216 \\
\midrule
 30\% & 10\% &21,751 & 145,009 & 299,080 & 1,531,654 & 3,072,371 & 15,398,105 & 30,805,273 \\
 30\% & 20\% &5,438 &  36,252 &  74,770 &   382,913 &   768,093 &  3,849,526 &  7,701,318 \\
 30\% & 30\% &2,417 &  16,112 &  33,231 &   170,184 &   341,375 &  1,710,901 &  3,422,808 \\
 30\% & 40\% &1,359 &   9,063 &  18,693 &    95,728 &   192,023 &    962,382 &  1,925,330 \\
  \midrule
 60\% & 10\% &11,064 & 79,905 & 165,957 & 854,372 & 1,714,890 & 8,599,037 & 17,204,221 \\
 60\% & 20\% &2,766 & 19,976 &  41,489 & 213,593 &   428,723 & 2,149,759 &  4,301,055 \\
 60\% & 30\% &1,229 &  8,878 &  18,440 &  94,930 &   190,543 &   955,449 &  1,911,580 \\
 60\% & 40\% &691 &  4,994 &  10,372 &  53,398 &   107,181 &   537,440 &  1,075,264 \\
  \midrule
 90\% & 10\% &4,553 & 37,946 & 79,686 & 413,607 & 831,009 & 4,170,221 & 8,344,237 \\
 90\% & 20\% &1,138 &  9,486 & 19,921 & 103,402 & 207,752 & 1,042,555 & 2,086,059 \\
 90\% & 30\% &506 &  4,216 &  8,854 &  45,956 &  92,334 &   463,358 &   927,137 \\
 90\% & 40\% &285 &  2,372 &  4,980 &  25,850 &  51,938 &   260,639 &   521,515 \\
  \bottomrule
\end{tabular}
\end{table}

In Figure \ref{fig:figure-4} the relationship between the sample size and the incidence rate looks linear on log-log scale as they have a power law relationship. However, while the two methods coincide at high incidence rates, pooled Wald method significantly underestimates the sample sizes at low incidence rates especially when the efficacy is high (note that y-axis is on logarithmic scale). Contrasting Figure \ref{fig:figure-4} with the case rates in Table \ref{tab:table-1}, it is clear that, to achieve the narrow confidence bounds that Pfizer and Moderna have reported, they would have needed several times more samples under pooled Wald method, and an order of magnitude more under Cramér–Rao bound. If the event rate were to differ from that in the general population or if possibility of misclassification was non negligible, such a discrepancy in incidence rates could cause such large variations in the variance that the trial population could be unrepresentative of the larger population. Table \ref{tab:table-2} provides the total sample sizes from Cramér–Rao bound formula \ref{eqn:ssizeCramer} for different levels of efficacy and effect size. It is clear that the sample size is also very sensitive to the choice of $\Delta$, therefore an investigator must be wary of misspecification of the anticipated treatment difference \cite{10.1093epirev}. 

Throughout the Methods, we incorporated the misclassification error in the calculations in order to emphasise the importance of accounting for classification bias when the disease is rare. We showed that, while lack of diagnostic sensitivity consistently inflates the estimated efficacy rates, imperfect specificity results is serious loss of accuracy and precision at low disease risks. Case definition for COVID-19 is particularly a major caveat. The three vaccine trials broadly follow FDA definition of the disease. For primary end points symptomatic cases are identified by surveillance or are self-reported, and are subsequently confirmed with RT-PCR. Pre-selecting of the participants for PCR assay could create the possibility for collider bias \cite{pmid33184277}. Moreover, the highly non-specific symptoms of COVID-19, which include symptoms as common as cough and congestion, could create the perfect conditions for misclassification. False negatives due to e.g. selective reporting, specimen collection, etc, and PCR false positives due to e.g. remnant viral RNA, etc could be introduced if the test is not repeated \cite{pmid33301459,balayla2020bayesian}. Much remains unknown about COVID-19 and its many symptoms and presentations. Therefore, it is recommended to account for classification bias in the calculation. The code for calculating the posterior probability of the vaccine efficacy, which can simultaneously marginalise over the diagnostic sensitivity and specificity is provided.


\section*{Code}
R code for the posterior probability of the efficacy was modified from code published in \cite{608719}. It is provided in Appendix along with functions to calculate the sample sizes from equations \ref{eqn:ssizeWald} and \ref{eqn:ssizeCramer}.

\section*{Acknowledgments}
The author's position at the University of Cambridge is funded by CRUK grant C60100/A23916. The author would like to appreciate the helpful comments received from the Cancer Mutagenesis group at MRC Cancer Unit.

\nolinenumbers

\bibliography{library}

\bibliographystyle{unsrt2authabbrvpp}

\section*{Appendix}
\begin{lstlisting}[breaklines=true]
#
# R function for Bayesian estimation of the efficacy rate using an
# imperfect test, originally modified from code provided by
# P. J. Diggle. Estimating prevalence using an imperfect test. \
# Epidemiology Research International, 2011(608719), 2011
#
# Notes
#
# 1. Prior for efficacy rate is assumed uniform on (0,1)
# 2. Priors for sensitivity and specificity are independent scaled
# beta distributions
# 3. Function uses a simple quadrature algorithm with number of
# quadrature points as an optional argument "ngrid" (see below);
# the default value ngrid=20 has been sufficient for all examples
# tried by the author, but is not guaranteed to give accurate
# results for all possible values of the other arguments.
#
library(Rmpfr)
.N <- function(.) mpfr(., precBits = 200)

efficacy.bayes<-function(alpha,Tc,n,pi,lowse=0.5,highse=1.0,
                           sea=1,seb=1,lowsp=0.5,highsp=1.0,
                           spa=1,spb=1,ngrid=20) {
  #
  # arguments
  # alpha: vector of efficacy rates for which posterior is required
  # (will be converted internally to increasing sequence of equally
  # spaced values, see "result" below)
  # Tc: number of positive test results
  # n: number of indiviudals tested
  # lowse: lower limit of prior for sensitivity
  # highse: upper limit of prior for sensitivity
  # sea,seb: parameters of scaled beta prior for sensitivity
  # lowsp: lower limit of prior for specificity
  # highsp: upper limit of prior for specificity
  # spa,spb: parameters of scaled beta prior for specificity
  # ngrid: number of grid-cells in each dimension for quadrature
  #
  # result is a list with components
  # alpha: vector of efficacy rate for which posterior density has
  # been calculated
  # post: vector of posterior densities
  # mode: posterior mode
  # interval: maximum a posteriori credible interval
  #
  ibeta<-function(x,a,b) {
    pbeta(x,a,b)*beta(.N(a),.N(b))
  }
  nalpha<-length(alpha)
  bin.width<-(alpha[nalpha]-alpha[1])/(nalpha-1)
  alpha<-alpha[1]+bin.width*(0:(nalpha-1))
  integrand<-array(0,c(nalpha,ngrid,ngrid))
  h1<-(highse-lowse)/ngrid
  h2<-(highsp-lowsp)/ngrid
  for (i in 1:ngrid) {
    se<-lowse+h1*(i-0.5)
    pse<-(1/(highse-lowse))*dbeta((se-lowse)/(highse-lowse),sea,seb)
    for (j in 1:ngrid) {
      sp<-lowsp+h2*(j-0.5)
      psp<-(1/(highsp-lowsp))*dbeta((sp-lowsp)/(highsp-lowsp),spa,spb)
      if(ngrid==1){
        se=highse; sp=highsp; pse=1; psp=1; h1=1; h2=1
      }
      c1<-1-sp
      c2<-se+sp-1
      g<-(c1+c2*pi)*chooseMpfr(n,Tc)*(ibeta(c1+c2*pi,Tc-1,n-Tc+1)-ibeta((c1+c2*pi)/2,Tc-1,n-Tc+1))
      p<-(c1+c2*pi)/(2-alpha)
      density<-rep(0,nalpha)
      for (k in 1:nalpha) {
        density[k]<-asNumeric(dbinom(Tc,n,.N(p[k]))/g) 
      }
      integrand[,i,j]<-pse*psp*density
    }
  }
  post<-rep(0,nalpha)
  for (i in 1:nalpha) {
    post[i]<-h1*h2*sum(integrand[i,,])
  }
  
  ord<-order(post,decreasing=Tc)
  mode<-alpha[ord[1]]
  
  cumpost=cumsum(post/sum(post))
  interval=c(alpha[which.min(abs(cumpost-0.025))],
             alpha[which.min(abs(cumpost-0.975))])
  
  list(alpha=alpha,post=post,mode=mode,interval=interval)
}
#
# example, to reproduce Table 1 results set ngrid=1, highse=1, highsp=1 
# oxford  30 of 5807 vs 101 of 5829  
# pfizer  8 of 18198 vs 162 of 18325      
# moderna 11 of 14134 vs 185 of 14073 

N<-14134+14073 # total number of participants
Tc<- 185 # number of cases in control
pi<-(185+11)/(14134+14073) # total number of cases / N 

ngrid<-1
lowse<-1
highse<- 1 #0.95
lowsp<-1
highsp<-1 #0.999
sea<-2 #5
seb<-2
spa<-2 #5
spb<-2
alpha<-seq(0,1,by=0.0005) #0.001*(1:400)
result<-efficacy.bayes(alpha,Tc,N,pi,lowse,highse,
                         sea,seb,lowsp,highsp,spa,spb,ngrid)
result$mode 
result$interval
plot(result$alpha,result$post/sum(result$post),type="l",xlab="alpha",ylab="p(alpha)")


# Function to reproduce per-group sample sizes presented in O'Neill, 
# 1988. Note that he used relative width (RW) and attack rate in 
# unvaccinated group (ARU)
sample_size_95wald_paper <- function(ARU,RW){
  VE=0.4
  y<-RW*VE/(2*(1-VE))
  d<-log(y+sqrt(y^2+1))
  return((1.96)^2/d^2*((1+1/(1-VE))/ARU-2))
}
print(t(outer(c(0.01,0.005,0.001,0.0005),rev(seq(0.1,1,0.1)),sample_size_95wald_paper)),quote = FALSE)

# Total sample sizes for pooled Wald variance (power 80% 
# and alpha=0.05)
sample_size_wald <- function(VE,Pi,delta){
  RW=delta/VE
  ARU=Pi/(2-VE)
  y<-RW*VE/(2*(1-VE))
  d<-log(y+sqrt(y^2+1))
  return(2*(1.96+0.84)^2/d^2*((1+1/(1-VE))/ARU-2))
}

# Total sample sizes for Cramer-Rao bound (power 80% and alpha=0.05)
sample_size_cramer <- function(VE,Pi,delta){
  return(4*(1.96+0.84)^2*(2-VE)^2*(2-VE-Pi)/Pi/delta^2)
}
\end{lstlisting}

\end{document}